\documentclass[a4paper,prl,twocolumn,superscriptaddress,showpacs]{revtex4}

\usepackage{amssymb}
\usepackage{amsbsy}
\usepackage{amsfonts}
\usepackage{bm}
\usepackage{graphicx}

\begin{document}

\title{Spectral diagrams of Hofstadter type for the Bloch electron in three dimensions}

\author{J.~Br\"uning}
\affiliation{Mathematisch-Naturwissenschaftliche Fakult{\"a}t II
der Humboldt-Universit\"at zu Berlin, Rudower Chaussee 25, Berlin
12489 Germany}
\author{V.~V.~Demidov}
\affiliation{Laboratory of Mathematical Physics, Mordovian State
University, Saransk 430000 Russia}
\author{V.~A.~Geyler}
\affiliation{Laboratory of Mathematical Physics, Mordovian State University,
Saransk 430000 Russia}

\begin{abstract}
Flux--energy and angle--energy diagrams for an exact
three-dimensional Hamiltonian of the Bloch electron in a uniform
magnetic field are analyzed. The dependence of the structure of
the diagrams on the direction of the field, the geometry of the
Bravais lattice and the number of atoms in an elementary cell is
considered. Numerical evidence is given that the angle--energy
diagram may have a fractal structure even in the case of a cubic
lattice. It is shown that neglecting coupling of Landau bands
changes considerably the shape of the diagrams.
\end{abstract}


\maketitle

The spectral properties of an electron in a two-dimensional (2D)
periodic structure with the Bravais lattice $\Lambda$ in the
presence of a uniform magnetic field $\mathbf{B}$ are determined
by commensurability or non-commensurability of two geometric
parameters: the area $S$ of an elementary cell of $\Lambda$ and
the square of the magnetic length $l_M^2=\hbar/\mu\omega$ (here
$\mu$ is the mass of the electron and $\omega$ is the cyclotron
frequency). If $S/2\pi l_M^2$ is a rational number, then the
electron energy spectrum has band structure, otherwise the
spectrum is a fractal set. As a result, the flux--energy diagram
for the spectrum has a remarkable recursive structure predicted by
M.~Ya.~Azbel' and numerically discovered by  D.~R.~Hofstadter in
the framework of the tight-binding approximation \cite{Hof}. The
essential point in the appearance of such a structure is the size
quantization of the electron motion along the field direction;
hence it seems likely that in three-dimensional (3D) periodic
systems, the fractal structure of the flux--energy diagram must
disappear \cite{HKZ}. Surprisingly, it was shown recently in a
series of papers  \cite{Kosh} that a fractal structure is visible
in the diagram depicting the dependence of the spectrum on the
angle between the field $\mathbf{B}$ (with fixed strength $B$) and
a fixed vector in $\Lambda$, if $\Lambda$ is an {\it anisotropic
rectangular} lattice. Moreover, a series of energy gaps as in
Hofstadter's butterfly arise in the {\it isotropic} case unless
$\mathbf{B}$ points in high-symmetry crystallographic directions
\cite{Kosh2}.

The tight-binding approximation method used in \cite{Hof},
\cite{Kosh}, \cite{Kosh2} is based on a series of considerable
simplifications of the initial periodic Landau operator, which
imposes severe limitations of the method. In particular, the
models considered in \cite{Kosh} and \cite{Kosh2} give no way to
take into account the effects of the interaction between Landau
levels, which have profound effects on the shape of flux--energy
diagrams and on the accompanying integer quantum Hall conductivity
\cite{KSSK}. Even without the consideration of the Landau band
coupling, the flux--energy diagrams obtained for the periodic
Landau operator differ considerably from those for the
tight-binding model \cite{Pets}. In this paper, we get rid of the
restrictions imposed by the tight-binding approximation method and
analyze the angle--energy and the flux--energy diagrams for the 3D
Landau operator perturbed by periodic potentials of various
geometries. To deal with an explicitly solvable model, we consider
periodic perturbations represented as sums of short-range
potentials; in the zero-range limit we get an explicit expression
for the dispersion relations, which is very useful for the
numerical analysis. Such potentials were already used in the
context of spectral problems for a 3D magneto-Bloch electron
\cite{GMA}, \cite{GD}. In the 2D case, the flux--energy diagrams
for zero--range potentials were considered in \cite{GrGe}.

The Hamiltonian $H$ being studied is the sum $H=H_0+V$, where
\begin{equation}
               \label{1}
H_0=\frac{1}{2\mu}\left({\bf p}-\frac{e}{c}{\bf A}({\bf
r})\right)^2
\end{equation}
is the Landau operator with the vector potential ${\bf A}({\bf
r})={\bf B}\times{\bf r}/2$, and $V$ is the potential of a crystal
lattice $\Gamma$ with the Bravais lattice $\Lambda$. Let ${\rm K}$
be the set of all nodes of $\Gamma$ lying in an elementary cell of
$\Lambda$, then $V$ is represented in the form
\begin{equation}
                     \label{poten1}
V({\bf r})= \sum\limits_{\bm{\kappa}\in\,{\rm
K}}\sum\limits_{\bm{\lambda}\in\Lambda}V_{\bm{\kappa}}({\bf
r}-\bm{\lambda})\,,
\end{equation}
where $V_{\bm{\kappa}}({\bf r})$ is the confinement potential of
the node $\mathbf{\kappa}$ which is supposed to be short-range.
More precisely, we choose $V_{\bm{\kappa}}({\bf r})=c_\kappa W(\bf
r-\bm{\kappa})$, where $W({\bf r})\sim 0$ outside a small sphere
of radius $R$ centered at zero, $\int W({\bf r})\,d{\bf r} =1$,
and the coupling constant $c_{\bm{\kappa}}$ is of order $R$. At
the zero-range limit, $R\to 0$, the potential $V_{\bm{\kappa}}$ is
characterized by one parameter only, namely, by the scattering
length $\rho_\kappa$, which is related to the binding energy of
the ground state for $V_{\bm{\kappa}}$ by
$E_{\bm{\kappa}}=-\hbar^2/2\mu\rho_{\bm{\kappa}}^2$ \cite{DA}.
Moreover, at this limit the Green function $G$ of $H$, $G({\bf
r},{\bf r}';E)=\langle{\bf r}|(E-H)^{-1}|{\bf r}'\rangle$, has the
following explicit expression in terms of the Green function
$G_0({\bf r},{\bf r}';E)$ of $H_0$ \cite{DA}, \cite{GM87}:
\begin{eqnarray}
                \nonumber
G({\bf r},{\bf r}';E)=G_0({\bf r},{\bf r}';E)
-\sum_{\bm{\gamma},\bm{\gamma}'\in\Gamma} G_0({\bf
r},{\bm{\gamma}};E)\\
               \label{5}
\times\left(Q^{-1}(E)\right)_{\bm{\gamma},\bm{\gamma}\,{}'}
G_0(\bm{\gamma}\,{}',{\bf r}';E)\,.
\end{eqnarray}
Here $Q^{-1}(E)$ is the matrix inverse to the infinite matrix
$Q(E)$ with elements
\begin{eqnarray}
               \nonumber
Q_{\bm{\gamma},\bm{\gamma}\,{}'}(E)&=&\left[G_0^{\rm
ren}(\bm{\gamma},\bm{\gamma};E)-
\frac{\mu}{2\pi\hbar^2\rho_{\bm{\gamma}}}
\right]\delta_{\bm{\gamma},\bm{\gamma}\,{}'}\\
               \label{6}
&&+(1-\delta_{\bm{\gamma},\bm{\gamma}\,{}'})G_0(\bm{\gamma},\bm{\gamma}\,{}';E)\,,
\end{eqnarray}
where $G_0^{\rm ren}$ denotes the renormalization of the
unperturbed Green function $G_0$:
$$
G_0^{\rm ren}({\bf r},{\bf r}';E)=G_0({\bf r},{\bf
r}';E)-\frac{\mu}{2\pi\hbar^2} \frac{\exp\left[-i\pi{\bf b}({\bf
r}\times{\bf r}')\right]} {|{\bf r}-{\bf r}'|}\,,
$$
with ${\mathbf b}=\mathbf{B}e/2\pi\hbar c$ (note that
$\phi_0=2\pi\hbar c/e$ is the magnetic flux quantum, therefore
$\mathbf{b}$ is the density of the magnetic flux). The explicit
form of $G_0$ is well known \cite{GM87}, \cite{Gou}:
\begin{equation}
          \label{Green}
G_0({\bf r},{\bf r}';E)=\Phi({\bf r},{\bf r}')F({\bf r}-{\bf
r}';E)\,,
\end{equation}
where
$$
\Phi({\bf r},{\bf r}')=\frac{\mu}{2^{3/2}\pi\hbar^2l_M}
\exp\left[-\pi i{\bf b}({\bf r}\times{\bf r}')- \frac{({\bf
r}_\bot-{\bf r}_\bot')^2}{4l_M^2}\right]\,,
$$
$$
F({\bf r};E)=\int\limits_0^\infty\frac{\exp\left[- \left({\bf
r}_\bot^2(e^t-1)^{-1}+{\bf r}_{||}^2t^{-1}\right)/2l_M^2\right]}
{\left(1-e^{-t}\right)\exp\left[(1/2-E/\hbar\omega)t\right]}
\frac{dt}{\sqrt{\pi t}}
$$
\begin{equation}
               \label{19}
=\sum_{l=0}^\infty\frac{\exp\left[-\sqrt{2(\varepsilon_l-E)/\hbar\omega}
|{\bf r}_{||}|/l_M\right]} {\sqrt{(\varepsilon_l-E)/\hbar\omega}}
L_l({\bf r}_\bot^2/2l_M^2)\,.
\end{equation}
Here ${\bf r}_{||}$ is the projection of ${\bf r}$ on the
direction of the field ${\bf B}$, ${\bf r}_{\bot}={\bf r}-{\bf
r}_{||}$, $L_l(x)$ denotes the $l$-th Laguerre polynomial, and
$\varepsilon_l=\hbar\omega(l+1/2)$. Note that $G_0^{\rm
ren}({\mathbf r},{\bf r}';E)$ is well defined at ${\mathbf r}={\bf
r}'$ and independent of ${\bf r}$:
\begin{equation}
                 \label{sGr}
G_0^{\rm ren}({\bf r},{\bf
r};E)=\frac{\mu}{2^{3/2}\pi\hbar^2l_M}\,\zeta\left({1\over2},{1\over2}
-{E\over\hbar\omega}\right)\,,
\end{equation}
where ${\zeta}(s,v)$ is the generalized Riemann (or Hurwitz)
$\zeta$-function \cite{BE}.

It is clear that at least for $E<\hbar\omega/2$ the energy $E$
belongs to the spectrum of $H$ if the matrix $Q(E)$ is not
invertible. Using the irreducible representations of the magnetic
translation group (MTG) for $H$ \cite{Zak}, the problem to invert
the infinite matrix $Q(E)$ may be reduced to a problem of
finite-dimensional algebra. This reduction requires the so-called
"rationality condition": the field $\mathbf{B}$ is said to be {\it
rational} with respect to the lattice $\Lambda$, if for a basis
${\mathbf a}_1$, ${\mathbf a}_2$, ${\mathbf a}_3$ of $\Lambda$,
the numbers $\mathbf{b}({\mathbf a}_j\times {\mathbf a}_k)$ are
rational \cite{Zak}. In this case a basis ${\mathbf a}_j$
($j=1,2,3$) can be chosen in such a way that $\eta\equiv
\mathbf{b}({\mathbf a}_1\times {\mathbf a}_2)>0$ and
$\mathbf{b}({\mathbf a}_2\times {\mathbf a}_3)=\mathbf{b}({\mathbf
a}_1\times {\mathbf a}_3)=0$. Let $\eta=N/M$, where $N$ and $M$
are coprime positive integers. Then all the irreducible
representations of the MTG which are trivial on the center of the
group are $M$-dimensional and are parameterized by a 3D torus
$\mathbf{T}$ \cite{Zak}. It is convenient to choose coordinates of
a point $\mathbf{p}$ from $\mathbf{T}$ such that $0\le
p_1<M^{-1}$, $0\le p_2,p_3<1$. Then we can form the following
$(MK)\times(MK)$ matrix $\widetilde Q$, where $K$ is the number of
nodes in ${\rm K}$:
\begin{eqnarray}
        \nonumber
\widetilde Q_{q,q'}({\bf p},E)=\exp[-\pi i m' {\bf
b}(\bm{\kappa}'\times {\bf a}_2)]\hfill\\
        \nonumber
\times\sum\limits_{\lambda_1, \lambda_2,
\lambda_3=-\infty}^{\infty} \exp\left[-\pi
i(2\bm{\lambda}\cdot{\bf p}+ \eta\lambda_1(M\lambda_2 + m ))\right]\\
        \nonumber
\times Q(\lambda_1{\bf a}_1 +(\lambda_2M + m){\bf a}_2 +
\bm{\kappa},\,m'{\bf a}_2
+\bm{\kappa}',\,\lambda_3{\bf a}_3;E)\\
         \label{KPR}
\times\exp\left[\pi i(\lambda_1{\bf a}_1 + (\lambda_2M + m){\bf
a}_2)({\bf b}\times\bm{\kappa})\right]\,.
\end{eqnarray}
Here $q$ denotes the pair $(m,\bm{\kappa})$ with
$\bm{\kappa}\in{\rm K}$, $m=0,\ldots,M-1$. Now the dispersion
relation for $H$ reads
\begin{equation}
      \label{KP}
  {\rm det}\,\widetilde Q({\bf p},E)=0\,.
\end{equation}

Equation (\ref{KP}) has for fixed $\mathbf{p}$ infinitely many
solutions $E_s(\mathbf{p})$ (dispersion laws), which are
continuous with respect to $\mathbf{p}$; each eigenvalue
$E_s(\mathbf{p})$ is $M$-fold degenerate. By definition, the
magnetic miniband $J_s$ is the set of all values of
$E_s(\mathbf{p})$. The minibands $J_s$ lying below $\hbar\omega/2$
form a piece of the spectrum of $H$ which may be attributed to
broadening the ground state of $H_0$; therefore, this piece is
nothing but the  {\it lowest Landau band}. According to
Eqs.~(\ref{KPR}) and (\ref{KP}), this band consists of $KM$
magnetic minibands which can overlap. If the overlapping is
absent, then the lowest Landau band approaches a Cantor set as
$\eta$ approaches an irrational number (and, therefore, $M$ tends
to infinity).

Using Eqs.~(\ref{KPR}) and (\ref{KP}) we analyze numerically the
structure of the lowest Landau band for various types of crystal
lattices and ranges of the field ${\bf B}$ employing two ways to
force $\eta$ to approach an irrational number: (1) we change the
value of $B$ keeping the direction of $\mathbf{B}$ fixed; (2) we
change the direction of $\mathbf{B}$ keeping the value of $B$
fixed. The lattice constant $a=7.5$~nm is chosen relevant to the
geometric parameters for the 3D regimented quantum dot
superlattice considered recently in \cite{Laz}. As to the
scattering length, we put $\rho\sim 1$~nm, this corresponds to the
binding energy $E\sim30$~meV.

\begin{figure}
\includegraphics[width=108pt]{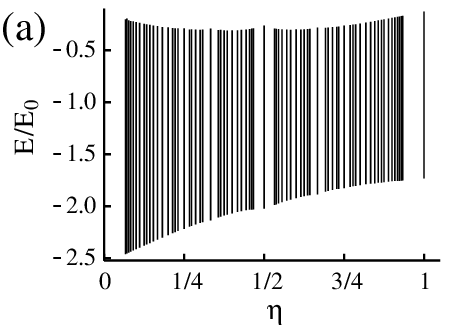}   
\includegraphics[width=75pt]{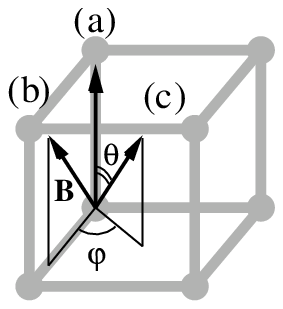}\\    
\includegraphics[width=196pt]{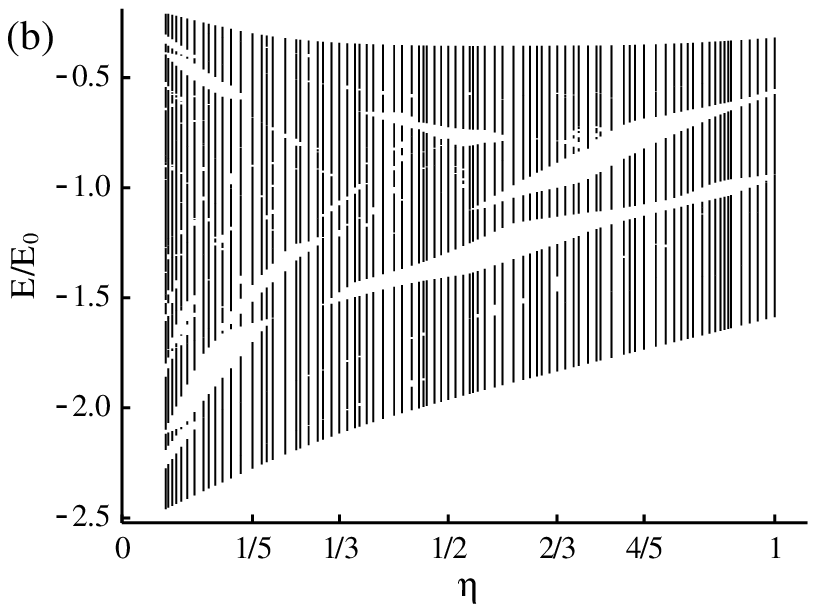}\\  
\includegraphics[width=196pt]{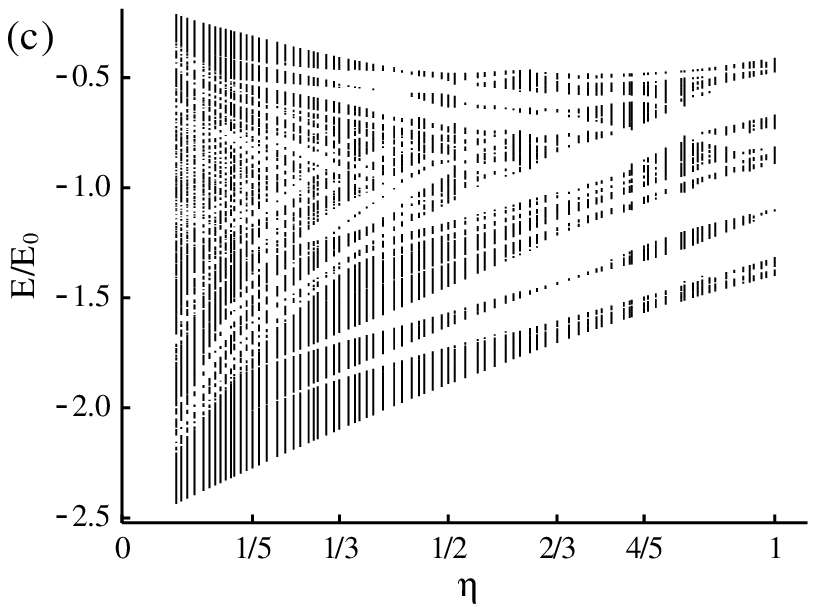}   
\caption{The lowest Landau band of the simple-cubic lattice
plotted against the magnetic flux for various directions of ${\bf
B}$: (a) ${\bf B}=(0,0,B)$; (b) ${\bf
B}=({3\over5}B,0,{4\over5}B)$; (c) ${\bf
B}=({12\over25}B,{9\over25}B,{4\over5}B)$. Here and below
$E_0=\hbar^2/\mu a^2$.}\label{fig1}
\end{figure}

\begin{figure}
\includegraphics[width=108pt]{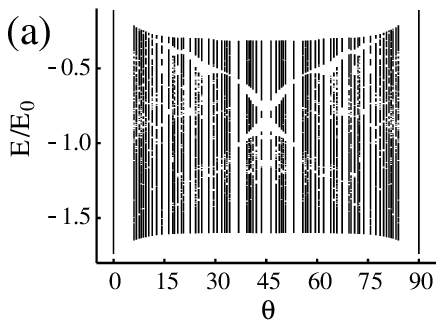}    
\includegraphics[width=75pt]{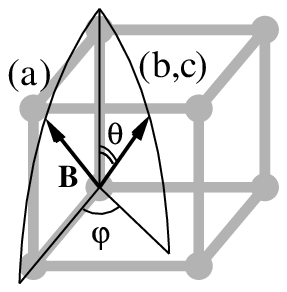}\\    
\includegraphics[width=196pt]{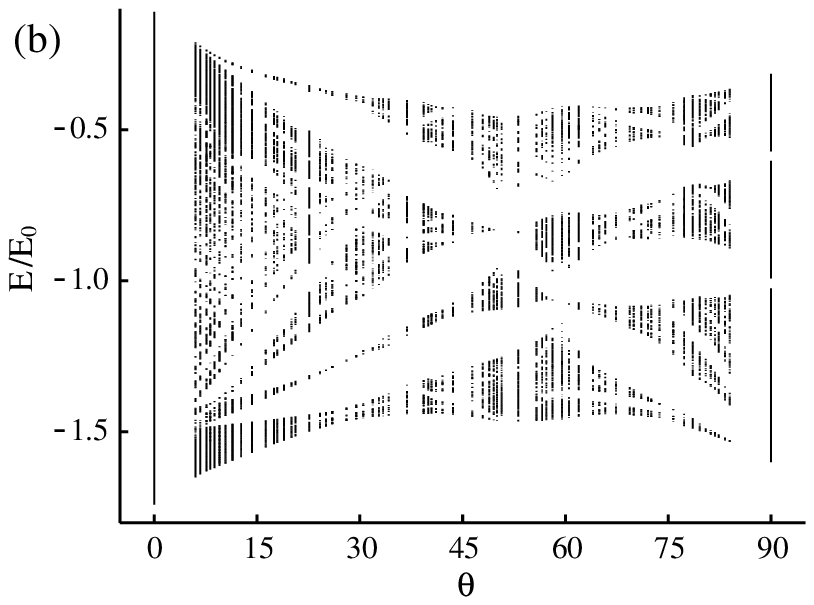}\\   
\includegraphics[width=196pt]{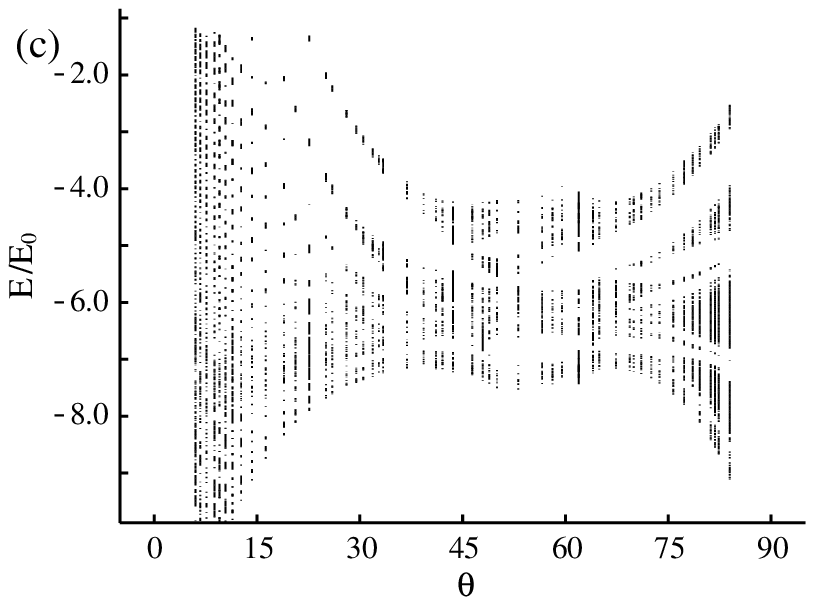}   
\caption{The lowest Landau band of the simple-cubic lattice
plotted against the tilting angle $\theta$ for various tilting
orientations of {\bf B}: (a) ${\bf B}=B(\sin\theta,0,\cos\theta)$;
(b,\,c) ${\bf
B}=B({4\over5}\sin\theta,{3\over5}\sin\theta,\cos\theta)$, in
panel (c) coupling between Landau levels is
neglected.}\label{fig2}
\end{figure}

\begin{figure}
\includegraphics[width=108pt]{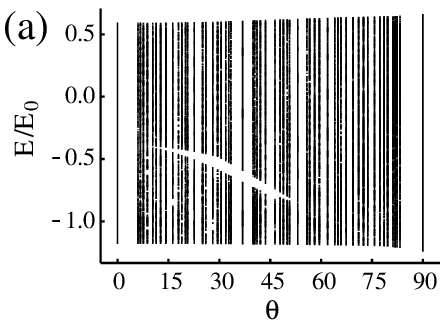}      
\includegraphics[width=75pt]{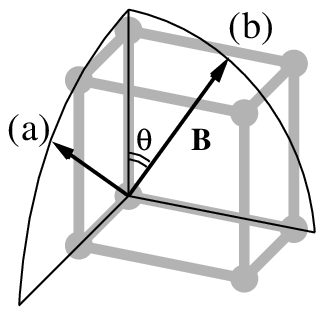}\\[2mm]  
\includegraphics[width=196pt]{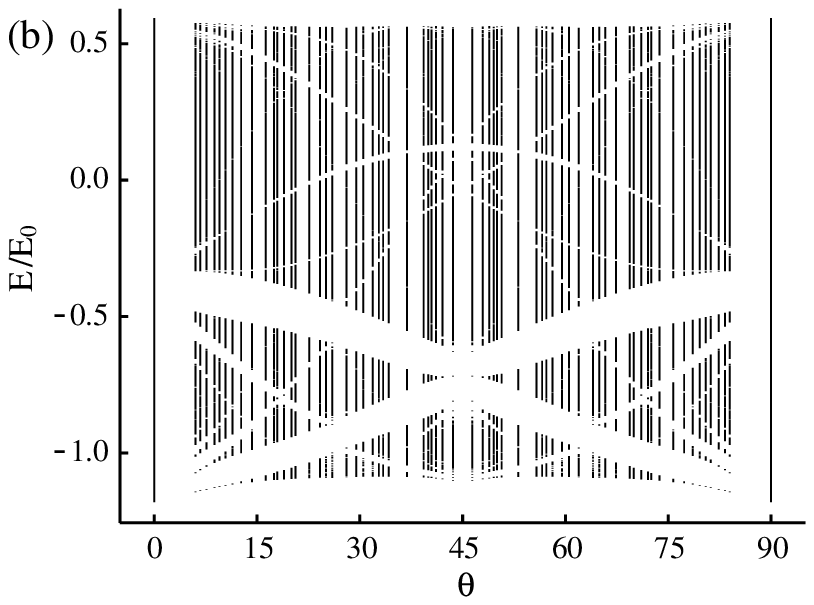}       
\caption{The lowest Landau band of the tetragonal monoatomic
lattice (with lattice constants $(a/2,a,a)$) plotted against the
tilting angle $\theta$ for various tilting orientations of {\bf
B}: (a) ${\bf B}=B(\sin\theta,0,\cos\theta)$; (b) ${\bf
B}=B(0,\sin\theta,\cos\theta)$.}\label{fig3}
\end{figure}

\begin{figure}
\includegraphics[width=108pt]{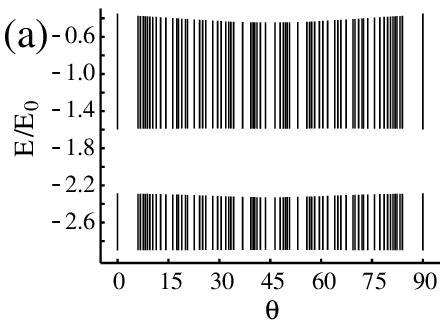}   
\includegraphics[width=75pt]{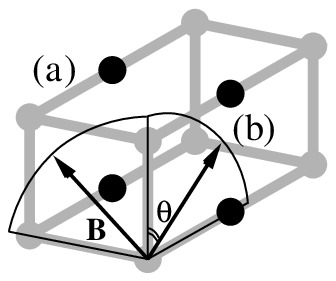}\\    
\includegraphics[width=196pt]{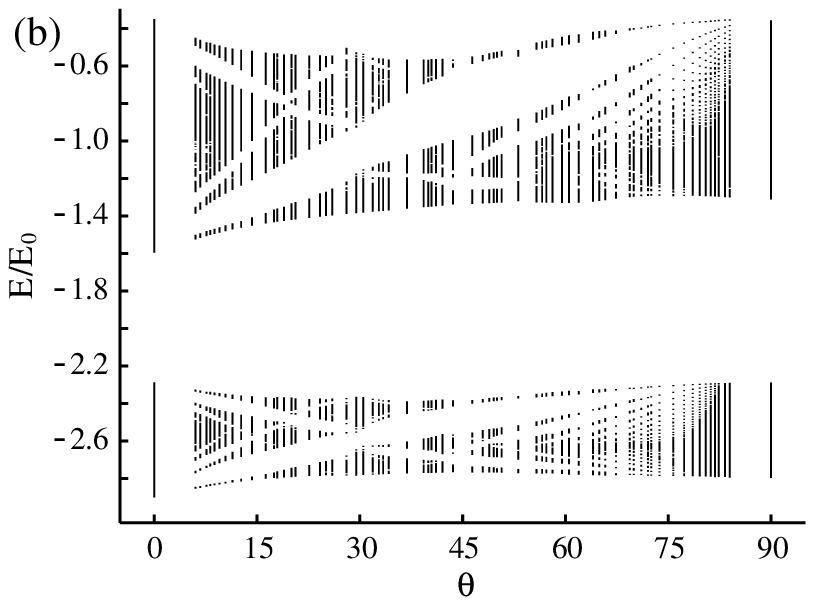}   
\caption{The lowest Landau band of the tetragonal double-atomic
lattice (with lattice constants $(2a,a,a)$) plotted against the
tilting angle $\theta$ for various tilting orientations of {\bf
B}: (a) ${\bf B}=B(0,\sin\theta,\cos\theta)$; (b) ${\bf
B}=B(\sin\theta,0,\cos\theta)$.}\label{fig4}
\end{figure}

The flux--energy diagrams for the simple-cubic lattice under
various directions of ${\bf B}$ are depicted in Fig.~1 (here
$\rho=2.5$~nm). If the magnetic field is directed along an edge of
the cubic elementary cell, then all subbands overlap and the gaps
are absent (Fig.~1a). If the symmetry of the magnetic field with
respect to the lattice decreases, then the number of open gaps
increases and has tendency to infinite magnification (Figs.~1b,c)
in full agreement with \cite{Kosh2}. Fig.~2 shows the
angle--energy diagrams for the lattice with the same geometric
parameters and $\rho=1.1$~nm (in Figs.~2--4 $\eta=1$; for
$a=7.5$~nm this corresponds to the field strength
$\thicksim70$~T). If the field ${\bf B}$ is rotated inside a face
of a cubic elementary cell, then the gaps, in general, overlap and
the diagram reveals no fractal structure (Fig.~2a). On the other
hand, if the rotation plane forms a dihedral angle $\varphi={\rm
atan}(3/4)$ with the plane of the face, then the fractal structure
of the diagram is clearly visible (Fig.~2b). Therefore, the
condition of anisotropy  \cite{Kosh} is not necessary for the
appearance of a fractal structure in the angle--energy diagram.
This can be understood taking the limit $B\to\infty$ in the matrix
$Q$. Eq.~(\ref{Green}) shows that in this limit $Q$ approximates
the matrix of the tight-binding Hamiltonian from \cite{Kosh} with
{\it energy-depending} coefficients $t_j$, and this dependence
leads to an anisotropy of the limiting matrix without any
anisotropy of the crystal lattice.

If in the representation (\ref{19}) we restrict ourselves only to
the first term, then this projection of the Green function on the
lowest Landau level models the system without interaction between
Landau levels. The corresponding diagram is given in Fig.~2c.

To compare our results to those from \cite{Kosh} we consider the
angle--energy diagrams for the tetragonal monoatomic lattice with
$\rho_0=2.5$~nm (Fig.~3). If the field $\mathbf{B}$ is rotated in
a face of an rectangular elementary cell, then we see a typical
1D-like energy spectrum in full agreement with \cite{Kosh}. It is
interesting that the {\it chemical anisotropy} radically
transforms the shape of the angle--energy diagram. To show this,
we consider a tetragonal double-atomic lattice with two sorts of
atoms in an elementary cell with scattering lengths
$\rho_1=2.5$~nm and $\rho_2=0.6$~nm (Fig.~4). If the magnetic
field is rotated in a face of anisotropy, then the angle--energy
diagram looks like deformed Hofstadter's butterfly. Since in this
case $K=2$, there is a doubling of minibands that causes the
appearance of a wide gap in the diagram ("atomic" gap). On the
other hand, if $\mathbf{B}$ rotates in the plane perpendicular to
the anisotropy axis, only the atomic gap appears in the diagram:
(Fig.~4a arises from Fig.~2a by the doubling of all bands).

As a conclusion, we have derived a new dispersion relation for a
3D Bloch electron in a uniform magnetic field (Eqs.~(\ref{KPR})
and (\ref{KP} )) and  analyzed the corresponding flux--energy and
angle--energy diagrams. For anisotropic rectangular crystals we
confirm the results of \cite{Kosh} and \cite{Kosh2} concerning the
structure of these diagrams. Moreover, we show that the anisotropy
is not necessary for the appearance of a multitude of gaps both in
flux--energy and angle--energy diagrams. Neglecting coupling of
Landau bands leads to a substantial deformation of angle--energy
diagram. Chemical anisotropy of the considered crystal (the
presence of distinct sorts of atoms in an elementary cell of the
Bravais lattice) leads to another deformation of the diagram, in
particular, to the appearance of an additional wide gap.

The work is partially supported by DFG SFB-288
"Differentialgeometrie und Quantenphysik" and by Grants of
DFG--RAS, INTAS, and RFBR.

\end{document}